\documentclass[preprint,aps,showpacs]{revtex4}

\usepackage{graphicx}
\usepackage{dcolumn}
\usepackage{amsmath}
\usepackage{bm}

\begin{document}

%\preprint{BohmTheory-paper}

\title[Short Title]{Two-Time Correlation Functions:
\\ Stochastic and Conventional Quantum Mechanics}

\author{L.~Feligioni}\email{lorenzo@bu.edu}
\affiliation{Physics Department, Boston University, Boston MA 02215 USA}
\author{O.~Panella}\email{orlando.panella@pg.infn.it}
\affiliation{Istituto Nazionale di Fisica Nucleare, Sezione di Perugia,
Via A. Pascoli, Perugia 06123, Italy}
\author{Y.N.~Srivastava}
\affiliation{Dipartimento di Fisica
dell'Universit\`a di Perugia, and INFN Sezione di Perugia, Via A. Pascoli, Perugia 06123, Italy}
\author{A.~Widom}
%\thanks{Partially supported by the Physics Department
%\& INFN at the University of Perugia
%while this work was in progress.}
\affiliation{Physics Department, Northeastern University,
Boston MA 02115 USA}
\date{October 15, 2005, -- {\sf quant-ph/0202045} --}
%\date{July 7, 2005}

\begin{abstract}
An investigation of two-time correlation functions  is reported
within the framework of (i) Stochastic Quantum Mechanics  and (ii)
conventional Heisenberg-Schr\"odinger Quantum Mechanics. The
spectral functions associated with the two-time electric dipole
correlation functions are worked out in detail for the case of the
hydrogen atom. While the single time averages are identical for
stochastic and conventional quantum mechanics, differences arise in
the two approaches for multiple time correlation functions.
\end{abstract}


\maketitle

\section{Introduction}

While most working physicists pay homage to the Copenhagen
interpretation of the Heisenberg\--Schr\"odin\-ger qu\-an\-tum
mechanics ({\sf QM}), many others seek a more causal
re-interpretation. One ambitious effort in this direction has its
origin in the works of David Bohm~\cite{bohm1,bohm2}. Bohm employed a
formalism for computing the paths for quantum mechanical particles
closely analogous to the method of Hamilton and Jacobi. Nevertheless
the Bohm approach~\cite{brendl,durr1,durr2},
also known as Stochastic Quantum Mechanics ({\sf SQM}), has been
thought to reproduce {\em in all instances} the same
probability distributions as does {\sf QM}.

The research concerning {\sf SQM} involves a considerable number of
authors  dealing with various aspects, even if not always within the
terms as originally proposed by Bohm~\cite{nelson}. Studies of extension of the
Bohm approach to the relativistic case are also
available~\cite{holl,hestenes1,hestenes2,fanchi1,fanchi2}.
Notwithstanding the relatively difficult nature of {\sf SQM}
computations, it is a generally accepted belief that, where a
comparison is possible, {\sf SQM} and  {\sf QM} would give the same
results. This is indeed the case for the average values of
observables at a fixed time.

However, in a previous work~\cite{reding}, examples were reported in
which {\sf SQM} produces results different from {\sf QM}. The
examples involved the two-time correlation functions of the electric
dipole moment components in the hydrogen atom. Within {\sf SQM}, an
explicit numerical calculation was performed~\cite{reding} yielding
the related spectral function (Fourier transform of the two-time
correlation function) for the hydrogen atom in the excited state
\begin{math} |nlm\rangle = |211\rangle \end{math}.
It was found to be quite different from the corresponding quantity
in {\sf QM}.

Our purpose is to study in more detail the comparison between
the frequency spectral functions in the two theories.
The general definition of two time correlation
functions (for a generic system described by a {\em time independent}
Hamiltonian \begin{math} H \end{math}) is
\begin{equation}
\Phi^{AB}(t,t') = \frac{1}{2}
\left< A(t) B(t') + B(t') A(t) \right>.
\label{2timescorr}
\end{equation}
The related noise spectral functions are defined as the
Fourier transform
\begin{equation}
S^{AB}(\omega) = \int_{-\infty}^{+\infty}
e^{i\omega t} \Phi^{AB}(t)\frac{dt}{ 2\pi }.
\label{spectralfunc}
\end{equation}
{\em Sum rules} for the functions \begin{math} S^{AB}(\omega ) \end{math}
are investigated in terms of their \begin{math} k^{th} \end{math}-order moments \begin{math} \gamma^{(k)}\end{math},
\begin{equation}
\gamma^{(k)}=\int_{-\infty}^{+\infty} \omega^k S^{AB}(\omega) d\omega .
\label{moments}
\end{equation}

Taking up (for definiteness) the case of the electric dipole moment
\begin{math} \bm{p}(t) = e\, \bm{r}(t) \end{math}
of the hydrogen atom, the following issues are addressed: (i) An
explicit analytic form is derived (in terms of modified Bessel
functions) within {\sf SQM} for the noise spectral function of the
dipole moments for the first few excited states, explicitly
\begin{math}
\left|nlm\right>=\left|211\right>,\ \left|322\right>,
\ \left|321\right>\ {\rm and}\ \left|311\right>
\end{math}.
General formulas are given for special combinations of
quantum numbers, e.g.
\begin{math}
\left|n,l,m\right>=\left|n,n-1,n-1\right>\ {\rm and}
\ \left|n,n-1,n-2\right>
\end{math}.
(ii) A general proof is given concerning the asymptotic behavior of
the spectral functions of the dipole moment fluctuations. As
\begin{math} \omega \to \infty \end{math}, the spectral functions
vanish with a power law for both {\sf QM} and  {\sf SQM}, but with
{\em different} exponents. (iii) The moments of the spectral
functions are investigated both for {\sf QM} and  {\sf SQM}, showing
explicitly the second order moment differences.

In Sec.~II a brief review of the Stochastic Quantum Mechanics is
provided and in Sec.~III the general notion of the two-time
correlation function is defined. The related spectral function is
also introduced. In Sec.~IV, an explicit calculation is reported for
the hydrogen atom two-time correlation function of the electric
dipole moment. The spectral function asymptotic behavior for large
frequency is discussed. Sec.~V contains a discussion of moment sum
rules, and in the concluding Sec.~VI the differences between {\sf
QM} and {\sf SQM} are further explored.

\section{The Stochastic Quantum Mechanics and Particle Trajectories}

In {\sf SQM} the wave function
\begin{math} \psi (\bm{r},t) \end{math}
entering into the Schr\"odinger equation,
\begin{equation}
i\hbar \frac{\partial \psi}{\partial t}= H\psi ,\qquad
H=-\frac{\hbar^2}{ 2\mu }\nabla^2 +V,
\end{equation}
%(where \begin{math} \Delta =div\ {\bf grad} \end{math}),
 is
conveniently written in the form
\begin{math} \psi = R e^{iS/\hbar } \end{math}.
\begin{math} R \end{math} and  \begin{math} S \end{math}
are real functions. {\sf SQM} then provides a causal interpretation
for the two resulting coupled differential equations thus obtained;
i.e.
\begin{subequations}
\begin{eqnarray}
\frac{\partial S}{\partial t}+\frac{|\bm{\nabla} S|^2}{2\mu }+
V-\left(\frac{\hbar ^2}{ 2\mu }\right)\frac{\nabla^2 R} {R} &=&0,
\label{eqmulti-a}\\
\frac{\partial (R^2)} {\partial t}+ \bm{\nabla} \cdot
\left(\frac{R^2 \bm{\nabla} S} {\mu }\right)&=&0 \label{eqmulti-b}.
\end{eqnarray}
\label{eqmulti}
\end{subequations}
The first of the above two equations is of the Hamilton-Jacobi form.
This is generalized by the presence of a new term which takes into account
the quantum effects via the \emph{quantum potential} contribution
\begin{equation}
Q(\bm{r},t)=- \left({\hbar^2\over 2 \mu}\right) {\nabla^2
R(\bm{r},t) \over R(\bm{r},t)}.
\end{equation}
\begin{math} S  \end{math} is the {\sf SQM} version of
the Hamilton principal function. The equations of motion are computed
from
\begin{equation}
\label{eqmoto} \dot{\bm{r}}(t)=\bm{v}(\bm{r}(t),t),
\end{equation}
\begin{equation}
\label{eqmotob}
 \bm{v}(\bm{r},t)= \frac{1}{\mu}\bm{\nabla} S( \bm{r},t) =
 \frac{\hbar}{\mu} \mathfrak{Im}(\psi^{-1} \nabla \psi )
\end{equation}
here
 \begin{math} \bm{v}(\bm{r},t) \end{math} is the velocity
of the particle that passes through the point
\begin{math} \bm{r} \end{math} at time \begin{math} t \end{math}.

Eq.~(\ref{eqmulti-b}) highlights the statistical character of the
theory and is interpreted as a continuity equation. As with the
classical statistical description, one introduces quantities which
account for the particle properties of an ensemble of identical
systems (same Hamiltonian, same quantum state, etc.) with
trajectories. If the initial distribution in configuration space
$\rho(\bm{r},t_0)$ is assumed to be given by
$\rho(\bm{r},t_0)=R^2(\bm{r},t_00)=|\psi(\bm{r},t_0)|^2 $ then the
distribution $\rho(\bm{r},t)$ satisfies the continuity equation
provided that $\rho(\bm{r},t)  = |\psi (\bm{r},t)|^2 $ \emph{at all
times}. This expresses the time invariance of the
configuration-space measure ``$\rho(\bm{r},t)\, d^3\, \bm{r}$". The
probability distribution in configuration space given by $\rho
=|\psi|^2$  is called the \emph{quantum equilibrium} distribution. A
system is then said to be in quantum equilibrium when its
configurations are randomly distributed according to the quantum
equilibrium distribution~\cite{brendl}. This is the so-called
\emph{quantum equilibrium hypotesis} (QEH): if a system is described
by the wave function $\psi$ then its configurations are distributed
according to $\rho=|\psi|^2$.

%represents the probability density to find a particle in
%\begin{math}\bm{r} \end{math} at time  \begin{math}t \end{math}
%and

%\begin{math}\bm{\nabla} S / \mu \end{math} is the velocity of
%the particle.
The fact that
\begin{math} R^2(\bm{r},t)=|\psi(\bm{r},t)|^2 \end{math}
is the probability density that the particle is at
\begin{math}\bm{r} \end{math} at time
\begin{math} t \end{math} holds true in {\sf SQM},
assures that one finds the same results as in {\sf QM}. The
probability density $|\psi(\bm{r},0)|^2$  gives information on the
initial conditions necessary for the quantum Hamilton-Jacobi theory
to be applied, thereby allowing the determination of particle
trajectories through Eqs.(\ref{eqmoto}).

In {\sf SQM}, the probability density at time \begin{math} t
\end{math} is related to particle trajectories. Closely analogous to
classical statistical mechanics, starting from the initial
distribution
\begin{math}
\rho(\bm{r}_0,t_{0})=|\psi(\bm{r}_0,t_{0})|^2
\end{math},
one has:
\begin{equation}
\label{rhot} \rho( \bm{r},t)= \int d^3 \bm{r}_0 \delta \left[\bm{r}_0-
\bm{r}(t,\bm{r}_0)\right] \rho(\bm{r}_0,0)=|\psi(\bm{r},t)|^2.
\end{equation}

 Averaging the initial position \begin{math}
\bm{r}_0 \end{math} then yields the same average values of the
corresponding operators in quantum mechanics.

Consider the Hermitian operator
\begin{math}
\hat{A}=\hat{A}(\hat{\bm{r}},\hat{\bm{p}})
\end{math}.
In the \begin{math} \bm{r} \end{math} representation, in a state
\begin{math}
\psi(\bm{r},t)=\langle \bm{r}|\psi (t)\rangle
\end{math},
one has
\begin{equation}
\langle \hat{A}\, \rangle_t= \int \psi^*(\bm{r},t)\hat{A}(\bm{r},-i
\hbar \bm{\nabla}) \psi (\bm{r},t) d^3 \bm{r}.
\end{equation}
That \begin{math} \hat{A} \end{math} is Hermitian, allows
the definition of a {\em local expectation value} which,
 when integrated
over all space, yields the average value
\begin{math}\langle \hat{A} \rangle \end{math}.
One defines
\begin{equation}
{\cal A}(\bm{r},t)={\Re }e \left({\psi^* (\bm{r},t)\hat{A} \, \psi
(\bm{r},t) \over
 \psi^* (\bm{r},t)  \psi(\bm{r},t)}\right),
\end{equation}
such that
\begin{equation}
\label{one_time_av} \langle \hat{A}\, \rangle_{t} =\int {\cal
A}(\bm{r},t) R^2(\bm{r},t)d^3 \bm{r}.
\end{equation}

\section{TWO-TIME CORRELATION FUNCTIONS}

In the previous section it was shown how the two theories ({\sf SQM}
and {\sf QM}) are completely equivalent if one considers  average
values of operators {\em at one fixed time}. As already anticipated
above, in order to distinguish the two theories, one needs to
consider {\it two-time} dependent quantities such as the correlation
functions defined in Eq.(\ref{2timescorr}). The averaging procedure
is specified in what follows for both {\sf QM} and {\sf SQM}.

\subsection{Stochastic Quantum Mechanics}
As discussed in the previous section, the Bohm approach is able to
reproduce the density distribution at time
\begin{math} t \end{math}
from the initial probability distribution
\begin{math}
\rho(\bm{r}_0,t_0)=|\psi(\bm{r}_0,t_0)|^2
\end{math}
at time \begin{math} t_0 \end{math} while still employing the notion
of trajectories (see Eq.~(\ref{rhot})).
For the time evolution of a general quantity
\begin{math} {\cal A} \end{math},
the quantum equilibrium hypothesis (QEH) gives (see
Eqs.(\ref{rhot} and \ref{one_time_av}):
\begin{equation}
\label{hui} \langle \hat{A} \, \rangle_t  = \int \rho
(\bm{r},t){\cal A}(\bm{r},t) d^3 \bm{r}
 =
\int \rho(\bm{r}_{0},0) {\cal A}\big(\bm{r}
(\bm{r}_{0},t),0\big)d^3\bm{r}_0.
\end{equation}

The average value of two operators depending on
different times is a generalization of Eq.(\ref{hui}).
Given any two dynamic variables,
\begin{math} {\cal A}(\bm{r},t)\end{math} and
\begin{math} {\cal B}(\bm{r},t)\end{math}, the
average value of their product, weighed over the initial
condition by the weight function
\begin{math} |\psi(\bm{r}_{0},0)|^2 =\rho(\bm{r}_0) \end{math},
is just the {\sf SQM} two-time correlation function\cite{balescu};
It is
\begin{equation}
\label{G(t-t')} \Phi^{AB}(t-t')= \int \rho(\bm{r}_{0}) {\cal
A}(\bm{r},t) {\cal B}(\bm{r},t^\prime ) d^3 \bm{r}_0 ,
\end{equation}
where
\begin{math}
{\cal B}(\bm{r},t^\prime )= {\cal B}\big(\bm{r}(t^\prime ,{\bf
r}_{0}),0\big)
\end{math}
and similarly
\begin{math}
{\cal A}(\bm{r},t)= {\cal A}\big(\bm{r}(t,\bm{r}_{0}),0\big)
\end{math}.

\subsection{Quantum Theory}
To simplify the notation, consider the case of
a time independent Hamiltonian with a discrete spectrum of
eigenvalues
\begin{equation}
H \left|N\right> = E_{N} \left|N\right> .
\end{equation}
In {\sf QM} the two-time correlation function for a given
state \begin{math}|N\rangle \end{math} is
\begin{equation}
\label{G_AB}
\Phi^{AB}_{N}(t-t')={1\over 2}
 \left< N \right| \hat{A}(t)\hat{B}(t') +
\hat{B}(t')\hat{A}(t) \left| N\right> ,
\end{equation}
where (in the Heisenberg representation)
\begin{equation}
\hat{A}(t)= e^{i\hat{H}t/\hbar } \hat{A} e^{-i\hat{H}t/\hbar }.
\end{equation}
Consider the special case \begin{math}\hat{A}=\hat{B}\end{math}.
Then Eq.~(\ref{G_AB}) reduces to
\begin{equation}
\label{QMgeneral1}
\Phi^{AA}_{N}(t-t')={1\over 2}
\left(\Phi^{AA}_{N+}(t-t')+\Phi^{AA}_{N-}(t-t')\right),
\end{equation}
where
\begin{equation}
\label{QMgeneral2}
\Phi^{AA}_{N\pm }(t-t') = {1\over 2}
\sum_{M} \left|\left<M\right|\hat{A}\left|N\right>\right|^2
e^{\mp i\omega_{MN}(t-t^\prime )}
\end{equation}
and
\begin{math} \hbar \omega_{MN}=(E_{M}-E_{N}) \end{math}.
The Hamiltonian being time independent is reflected
by the fact that the \begin{math} \Phi_N^{A,B} \end{math}
depend just on \begin{math}(t-t')\end{math}.

The related noise spectral functions, defined as in
Eq.(\ref{spectralfunc}) in {\sf QM} are easily found to
be given by
\begin{equation}
S_N(\omega) = S_N^+(\omega) + S_N^-(\omega),
\label{QMspectrala}
\end{equation}
with
\begin{equation}
S_N^{\pm }(\omega)={1\over 2} \sum_M
\left| \left< M \right|\hat{A}\left| N \right> \right|^2
\delta ( \omega \mp \omega_{MN}).
\label{QMspectralb}
\end{equation}
When \begin{math}|N\rangle \end{math} is the ground state only
\begin{math} S_N^+ \end{math} contributes for \begin{math} \omega
\ge 0 \end{math}. In general, \begin{math} S_N(\omega ) \end{math}
is an even function of \begin{math}\omega \end{math}. Furthermore,
if the Hamiltonian has a spectrum with both a discrete  and a
continuous part, then the sums in the above
Eqs.(\ref{QMspectrala},\ref{QMspectralb}) clearly split into a
corresponding sum over the discrete part of the spectrum plus an
integral over the continuous one.
%Therefore, differences between the
%{\sf SQM} and {\sf QM} description may well be anticipated. While in
%Eqs.(\ref{QMgeneral1}) and (\ref{QMgeneral2}) all intermediate
%states give a contribution, in Eq.(14) only a given state
%\begin{math} |N\rangle \end{math} enters into the description of the system.
It will prove  useful in deriving the asymptotic behavior
of the {\sf QM} noise spectral function
\begin{math}S_N(\omega) \end{math} (in the next section)
to relate it to  the Fourier transform of the average of the
time-ordered product; i.e.
\begin{equation}
\alpha_N(\omega ) = {i\over \hbar } \int_{-\infty}^{+\infty}
e^{i\omega t} \left< N \right|T\left[\hat{A}(t) \hat{A}(0)\right]
\left|N \right>dt, \label{propagator}
\end{equation}
where ``$T$'' denotes operator time ordering. The quantities
\begin{math}S_N(\omega ) \end{math} and \begin{math} \alpha_N(\omega
) \end{math} are related by
\begin{equation}
S_N(\omega ) = {\hbar \over 2\pi } {Im } \ \alpha_N(\omega ).
\label{connection}
\end{equation}

\section{The electric dipole moment and  the hydrogen atom}

\subsection{Stochastic Quantum Mechanics}
Consider the electric dipole moment along the $x$ axis
\begin{math} d_x(t)=e x(t)\end{math}.
In the Bohm description,
the two-time correlation function of the electric dipole moment
for a generic {\em excited} eigenstate of the hydrogen
atom with quantum numbers \begin{math} (n,l,m) \end{math}
is as given in Eq.~(\ref{G(t-t')}); i.e.
\begin{eqnarray}
\Phi_{nlm}(t,t')&=&
\left<d_{x}(t)d_{x}(t')\right>_{nlm} \nonumber \\
&=&\int d_{x}\big(\bm{r}(t,\bm{r}_0 )\big) d_{x}\big(\bm{r}(t^\prime
,\bm{r}_0 )\big) \rho_{nlm}(\bm{r}_0)d^3\bm{r}_0, \label{SQMcorr}
\end{eqnarray}
where
\begin{math}
d_{x}(\bm{r}(t,\bm{r}_{0}))= e r_{0} \sin \theta_{0} \cos \phi(t)
\end{math}. Using the hydrogen atom bound state wave functions
and the solution to the equations of motion in the Bohm description,
given in appendix A, it is straightforward to derive the following
properties of the related noise spectral functions
\begin{eqnarray}
S_{n,l,m} (\omega) &=& S_{n,l,m} (-\omega), \nonumber \\
S_{n,l,m} (\omega) &=& S_{n,l,-m} (\omega), \nonumber \\
S_{n,l,0} (\omega) &\propto& \delta(\omega).
\end{eqnarray}
Thus, without loss of generality, \begin{math} m>0 \end{math} and
\begin{math} \omega >0 \end{math}
 can be  assumed in the following discussion.
For general quantum numbers, one can derive:
\begin{equation}
S_{nlm}(\omega) = {c_{nml}\over 128}\left({e^2 a_0^2\over
\omega_0}\right) [z_{n,m}(\omega )]^{2(3+m)} \, \int_{z_{n,m}(\omega
)}^\infty \negthinspace\negthinspace\negthinspace e^{-\rho}
\rho^{2(l-m)}[L_{n+l}^{2l+1}(\rho)]^2
\left({[C_{l-m}^{(m+1/2)}(\xi)]^2 \over \xi }\right) d\rho ,
\label{SQMgeneral}
\end{equation}
where
\begin{eqnarray}
\omega_{0}&=&(\hbar / \mu a_{0}^2),
\nonumber\\
z_{n,m}(\omega ) &=& {2\over n}\sqrt{m\omega_0\over \omega },
\nonumber\\
\xi &=& \sqrt{1-\left({z_{n,m}(\omega )\over \rho }\right)^2}
\end{eqnarray}
\begin{equation}
c_{nlm}=\left[{n^4(2l+1)(l-m)!((2m-1)!!)^2(n-l-1)!
\over 2nm (l+m)![(n+l)!]^3}\right],
\end{equation}
and \begin{math} C_{l-m}^{(m+1/2)}(\xi) \end{math}
are the ultra-spherical Gegenbauer
po\-ly\-no\-mials\cite{abramowitz} which satisfy the following relations:
\begin{eqnarray}
C_{l-m}^{(m+1/2)}(1)&=& {(l+m)! \over (2m)!(l-m)!}\nonumber \\
C_{0}^{(m+1/2)}(\xi)&=& 1 \nonumber \\
{C_{l-m}^{(m+1/2)}(\xi)\over 1\cdot 3 \cdots (2m-1)} &=&
\left( {d \over d \xi }\right)^mP_l(\xi) ,
\end{eqnarray}
where \begin{math} P_l(\xi ) \end{math} is the Legendre
polynomial.

A few special cases have been explicitly computed and are here
reported using \begin{math}{\cal S}_{nlm}(\omega)\end{math} to
denote the noise spectral function of Eq.(\ref{SQMgeneral}) in units
of
\begin{math}(e^2 a_{0}^2)/\omega_{0}\end{math} :
%In Fig.1 we show plots of some selected spectral functions
%\begin{math} \sigma_{nlm}(\omega ) \end{math} as defined by
\[
S_{nlm}^{{\sf SQM}}(\omega)= \left({e^2a_0^2\over
\omega_0}\right){\cal S}_{nlm}(\omega ),
\]

\noindent \underline{$\bm{n=2}$}
\begin{equation}
{\cal S}_{211}(\omega ) = \frac{1}{128}\left(\frac
{\omega_{0}}{\omega}\right)^4 \, z_{2,1}K_{1}(z_{2,1}) \label{S211}
\end{equation}

\noindent\underline{$\bm{n=3}$}
\begin{eqnarray}
\label{S322} {\cal S}_{322}(\omega) & = & \left({1\over
2187}\right)\left({\omega_{0}\over \omega}\right)^5 z_{3,2}K_{1}(z_{3,2})\cr
{\cal S}_{321}(\omega) & = &
\left({1\over 3888}\right) \left({\omega_{0}\over
\omega}\right)^4z_{3,1} \, \left[ 2\, K_{1}(z_{3,1}) + z_{3,1}\, K_{0}(z_{3,1})\right]\cr
{\cal S}_{311}(\omega) & = &
\left( {1\over 243}\right) \left({\omega_{0}\over \omega}\right)^4
z_{3,1}\, \left[\left({5\over 8}+{z_{3,1}^2\over
16}\right)K_1(z_{3,1})
-{7\over 16}z_{3,1}K_0(z_{3,1})\right]
\end{eqnarray}

\noindent \underline{$ \bm{n \ \ \ (l=m=n-1)} $}
\begin{eqnarray}
\label{snuno} {\cal S}_{n,n-1,n-1}(\omega) &=& c_n
\left({\omega_0 \over \omega}\right)^{n+2}
z_{n,n-1}K_1(z_{n,n-1}) \nonumber \\
c_n &=& {1\over 8}\left({2\over n}\right)^{2n}(n-1)^{n+1}
{\left[(2n-3)!!\right]^2 \over 2n[(2n-2)!]^2}
\end{eqnarray}

\noindent \underline{$ \bm{ n \ \ \  (l=n-1,\  m=n-2) } $}
\begin{eqnarray}
\label{sndue} {\cal S}_{n,n-1,n-2}(\omega) &=& {\bar{c}_n}
\left({\omega_0\over \omega}\right)^{n+1}
z_{n,n-2}^2 K_2(z_{n,n-2}) \nonumber \\
\bar{c}_n &=& c_{n,n-1,n-2}\left[(2n-1)!(2n-3)!!\right]^2
{4(n-2)\over 128n^2}
\end{eqnarray}
In the above expressions, \begin{math} K_n(z) \end{math} are the
\begin{math} n^{th} \end{math} order modified
Bessel functions.
%Fig.\ref{fig1} shows the above spectral functions
%as a function of \begin{math}(\omega/\omega_0)\end{math}.
It should be noted that the function \begin{math}S_{211}(\omega)\end{math}
agrees exactly with the numerical computation previously
reported~\cite{reding}.
It has been verified that the particular results of
Eqs.(\ref{S211}) and (\ref{S322}) are consistent with general formulas
given in Eqs.~(\ref{snuno}) and (\ref{sndue}). In Fig.1 we show the plots
of some of the above explicit examples.

\begin{figure}[h]
\begin{center}
\scalebox
{0.9}{\includegraphics*[109,405][500,720]{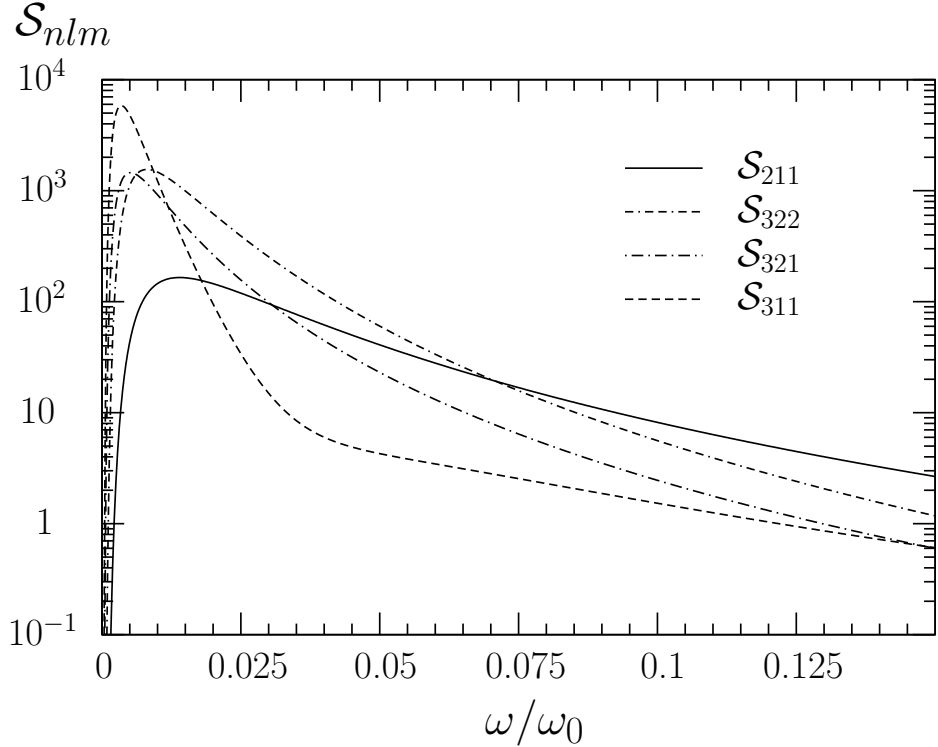}}\caption{Shown
are four plots of typical spectral functions ${\cal S}_{nlm}(\omega
)=\omega_0 S^{\sf SQM}_{nlm}(\omega )/(ea_0)^2$ for the electric
dipole moment of the Hydrogen atom computed employing the Stochastic
Quantum Mechanics.}
\end{center}
\label{fig1}
\end{figure}

\noindent{\bf Asymptotic behavior.}\\
From Eq.(\ref{SQMgeneral}) it is possible to derive {\em for general
quantum numbers} the asymptotic behavior at large frequencies of the
{\sf SQM} noise spectral function; it is
\begin{equation}
S_{nlm}^{{\sf SQM}}(\omega) \to {\cal C}_{nlm} \left({e^2 a_{0}^2
\over \omega_{0}}\right) \left(\frac
{\omega_{0}}{\omega}\right)^{3+m} \ {\rm as} \ \omega \to \infty ,
\label{asySQM}
\end{equation}
where
\begin{equation}
{\cal C}_{nlm} = {c_{nlm}\over
128}\left[C_{l-m}^{(m+1/2)}(1)\right]^2 \left({4m\over
n^2}\right)^{3+m} \, \int_0^\infty e^{-\rho }\rho^{2(l-m)}
\left[L_{n+l}^{(2l+1)}(\rho)\right]^2 d\rho .
\end{equation}

Thus, noise spectral functions vanish as
\begin{math} \omega \to \infty \end{math}
with a power law. The exponent is related to the state's quantum
numbers by \begin{math}(3 + m)\end{math}. The explicit cases
considered above can easily be shown to agree with
Eq.(\ref{asySQM}). One may use the expansion of the modified Bessel
functions for small values of the argument\cite{abramowitz}, i.e. as
\begin{math} z\to 0  \end{math}
\begin{eqnarray}
K_\nu (z) \to \left\{
\begin{array}{ll}
\displaystyle{(1/2)}\Gamma(\nu)
(\displaystyle{z/2})^{-\nu}
\ {\rm if}\ \nu \neq 0,\\
-\ln(z) \ \ \ \ \ \ \ \ \ \ \ \ \ \ \ {\rm if}
\ \nu = 0.
\end{array} \right.
\end{eqnarray}
One then infers the large
\begin{math} \omega \to \infty  \end{math}
behavior of the spectral functions; i.e.
\begin{eqnarray}
S_{nlm}^{\sf SQM}(\omega) \to \left\{
\begin{array}{ll}
\displaystyle{(1/128)}\left(\displaystyle{\omega_{0}\over \omega}\right)^4
\ \ \ {\rm for\ state }\ \left|211\right>, \\
\displaystyle{(1/2187)}
\left(\displaystyle\frac{\omega_{0}}{\omega}\right)^5
\ \ {\rm for\ state }\ \left|322\right>, \\
\displaystyle{(1/ 1944)}
\left(\displaystyle\frac{\omega_{0}}{\omega}\right)^4
\ \ {\rm for\ state }\ \left|321\right>, \\
\displaystyle{(5/1944)}
\left(\displaystyle\frac{\omega_{0}}{\omega}\right)^4
\ \ {\rm for\ state }\ \left|311\right>. \\
\end{array}\right.
\end{eqnarray}

\subsection{Quantum Mechanics}

As anticipated at the end of Sec.III, in order to derive the asymptotic
behavior of the {\sf QM} noise spectral function it proves useful to work
with the ``time ordered propagator'' defined in Eq.(\ref{propagator})
and use Eq.(\ref{connection}) to find \begin{math} S_N(\omega) \end{math}.
The function \begin{math} \alpha_N(\omega) \end{math}
can easily be connected with the (retarded) Green's function of the
Schr\"odinger equation; i.e.
\begin{equation}
G_{ret}(\bm{r}_2,\bm{r}_1;\omega) = -{i\over \hbar} \int_0^\infty
\left< \bm{r}_2\right| e^{i\{\omega -(H/\hbar )\}t}\left|{\bf
r}_1\right>dt. \label{retarded}
\end{equation}
The time ordered response function
\begin{math} \alpha_N^{(AB)}(\omega) \end{math}
is related to the Green's function
\begin{math}  G_{ret}(\bm{r}_2,\bm{r}_1;\omega) \end{math} via
\begin{eqnarray}
-\alpha_N^{(AB)}(\omega) &=& \int d^3\bm{r}_1 \int d^3\bm{r}_2
\left<N\right|A(0)\left|\bm{r}_2\right>\,
G_{ret}\left(\bm{r}_2,\bm{r}_1;{E_N\over \hbar}+\omega\right)
\left<\bm{r}_1 \right|B(0)\left|N\right> + \nonumber \\
&&\int d^3\bm{r}_1 \int d^3\bm{r}_2
\left<N\right|B(0)\left|\bm{r}_2\right>\,
G_{ret}\left(\bm{r}_2,\bm{r}_1;{E_N\over \hbar}-\omega\right)
\left<\bm{r}_1 \right|A(0)\left|N\right>.\nonumber\\
\label{central}
\end{eqnarray}
The retarded Green's function associated to the Hamiltonian
\begin{math} H \end{math}, as defined in Eq.(\ref{retarded}),
satisfies the differential equation
\begin{equation}
(\hbar\omega - H_{(\bm{r}_2)})G_{ret}(\bm{r}_2,\bm{r}_1;\omega )
=\delta(\bm{r}_2-\bm{r}_1),
\end{equation}
where
\begin{math} H_{(\bm{r}_2)} = -(\hbar^2/2\mu)\nabla^2_{(\bm{r}_2)} -(Ze^2/r_2)\end{math}
denotes a one electron Coulomb system. The differential equation of
the {\em non-relativistic} Coulomb Green's function (in the standard
normalization) is~\cite{mapleton,hostler,mano}
\begin{equation}
\left\{ \nabla^2_{(\bm{r}_2)}+\left({2k\nu \over r_2}\right) +k^2
\right\} G(\bm{r}_2,\bm{r}_1;\omega )= \delta(\bm{r}_2-\bm{r}_1),
\label{coulomb}
\end{equation}
where
\begin{math} k=\sqrt{(2\mu\omega/\hbar)}\end{math}
and
\begin{math}\nu=4\pi\hbar^2/(Z\mu e^2 k)\end{math}.

The Green's functions appearing in Eqs.(\ref{retarded}) and
(\ref{coulomb}) are  related  by a normalization constant
\begin{math}
G_{ret}(\bm{r}_2,\bm{r}_1;\omega ) = (2\mu/\hbar^2)
G(\bm{r}_2,\bm{r}_1;\omega )
\end{math}.
A closed expression  of the  Coulomb Green's function in terms of
Whittaker functions~\cite{abramowitz} has been given by L.
Hostler~\cite{hostler} as
\begin{equation}
G(\bm{r}_2,\bm{r}_1;\omega ) = - {\Gamma(1-i\nu)\over 4\pi|\bm{r}_2
- \bm{r}_1|} \, det
\begin{pmatrix}
{\cal W}_{i\nu;1/2}(-ik\alpha_2) & {\cal M}_{i\nu;1/2}(-ik\alpha_1) \\
\dot{\cal W}_{i\nu;1/2}(-ik\alpha_1) & \dot{\cal M}_{i\nu;1/2}(-ik\alpha_2)
\end{pmatrix}
\label{exactGreen}
\end{equation}
where the dots over the Whittaker functions denote differentiation
with respect to their arguments and
\begin{eqnarray}
\alpha_2&=& r_2 +r_1 + |\bm{r}_2 - \bm{r}_1|\nonumber \\
\alpha_1&=& r_2 +r_1 - |\bm{r}_2 - \bm{r}_1|.
\end{eqnarray}
\noindent{\bf Asymptotic behavior.}\\
The aim of this section is to derive the asymptotic form of
the noise spectral functions
as \begin{math}\omega \to \infty \end{math}.
Using Eq.(\ref{central}) requires the
Coulomb Green's function in the  regime
\begin{math}\omega \to \infty \end{math}
(respectively \begin{math}\omega \to -\infty \end{math}).
In this limit,
\begin{math} k\to \infty \end{math}
(respectively \begin{math}k \to i\infty \end{math}). Also
\begin{math} \nu \to 0 \end{math} so that Eq.(\ref{coulomb}) reduces
to the differential equation of the free particle Green's function
\begin{math} G_0( \bm{r}_2,\bm{r}_1;\omega)  \end{math}. Indeed,
from the exact solution in Eq.(\ref{exactGreen}), it is possible to
show explicitly that as \begin{math} |\omega| \to \infty  \end{math}
\begin{equation}
G\left(\bm{r}_2,\bm{r}_1;\omega\right) \to G_{0}\left(\bm{r}_2,{\bf
r}_1;\omega\right)= - {e^{ik|\bm{r}_2-\bm{r}_1|}\over 4\pi|\bm{r}_2
- \bm{r}_1|}.
\end{equation}
This proves that the {\em exact} Coulomb Green's function has
an oscillatory behavior at large positive frequencies
and an exponentially damped behavior at
large negative frequencies:
\begin{eqnarray}
-4\pi |\bm{r}_2-\bm{r}_1|G & \to &  e^{i|\bm{r}_2-{\bf
r}_1|\sqrt{2\mu \omega /\hbar }}
\ \ \ \ \ {\rm as}\ \omega\to +\infty , \nonumber \\
-4\pi |\bm{r}_2-\bm{r}_1|G & \to & e^{-|\bm{r}_2-{\bf
r}_1|\sqrt{2\mu |\omega| /\hbar }} \ {\rm as}\ \omega\to -\infty .
\end{eqnarray}
Inserting this result into the central Eq.(\ref{central}), one finds
that only the first term survives; i.e. for
\begin{math} \omega \to \infty  \end{math}
\begin{equation}
(\hbar^2/2\pi \mu)\alpha_N^{(AB)}(\omega)\to \int d^3\bm{r}_1\int
d^3\bm{r}_2 \left<N\right|A(0)\left|\bm{r}_2\right> \,
{e^{ik|\bm{r}_2-\bm{r}_1|} \over |\bm{r}_2-\bm{r}_1|} \, \left<{\bf
r}_2\right|B(0)\left|N\right>.
\end{equation}
where it is to be recalled that
\begin{math} k=\sqrt{(2\mu \omega /\hbar)}\end{math}.

The above considerations can be readily applied to the case
of the electric dipole moment in the hydrogen atom
with \begin{math} A(0)=B(0)=e x \end{math}. As
\begin{math} \omega \to \infty  \end{math}
\begin{equation}
\alpha_N^{(xx)}(\omega)\to  {\mu e^2 \over 2\pi \hbar^2} \int
d^3\bm{r}_1\,d^3\bm{r}_2 \, \psi_N^*(\bm{r}_2) \,x_2\,
{e^{ik|\bm{r}_2 - \bm{r}_1|} \over |\bm{r}_2 - {\bf r}_1|} \, x_1\,
\psi_N(\bm{r}_1) .
\end{equation}
Using the above in Eq.(\ref{connection}), the
asymptotic expression for the electric dipole moment
spectral function is derived; i.e.
\begin{equation}
S_N^{(xx)}(\omega)\to {e^2\over 4\pi ^2 a_0^2 \omega_0}  {Im}\, \int
d^3\bm{r}_1\int d^3\bm{r}_2\, \psi_N^*(\bm{r}_2)\, x_2\,
{e^{ik|\bm{r}_2 - \bm{r}_1|} \over |\bm{r}_2 - \bm{r}_1|}\,  x_1\,
\psi_N(\bm{r}_1).
\end{equation}
With \begin{math}\left|N\right>=\left|nlm\right> \end{math},
the hydrogen atom wave functions are written as
\begin{math}
\psi_{nlm}(\bm{r})=\chi_{nl}(r) Y_{lm}(\theta,\phi)
\end{math}. One may also employ the expansion
\begin{equation}
{e^{ik|\bm{r}_2 - \bm{r}_1|} \over |\bm{r}_2 - \bm{r}_1|} = (4\pi
ik) \sum_{l=0}^\infty j_l(kr_<) h^{(1)}_l(kr_>) \, \sum_{m=-l}^l
Y_{lm}^*(\theta_1,\phi_1) Y_{lm}(\theta_2,\phi_2),
\end{equation}
where
\begin{math}r_< = min(r_1,r_2)\end{math} and
\begin{math} r_> = max(r_1,r_2) \end{math}.
The noise spectral function then has the asymptotic limit
\begin{equation}
S_{nlm}^{(xx)}(\omega) \to {e^2\over \pi \omega_0a_0^2} {Im}\,
\Big\{ ik \sum_{l^\prime =0}^\infty C_{lm}^{l^\prime }
\int_0^\infty\int_0^\infty dr_1dr_2 \,
%\int_0^\infty
(r_2r_1)^3 \chi_{nl}(r_2)j_{l^\prime }(kr_<) h^{(1)}_{l^\prime
}(kr_>)\chi_{nl}(r_1) \Big\}, \label{LAST}
\end{equation}
with the constants \begin{math} C_{lm}^{(l')} \end{math} defined as
\begin{equation}
C_{lm}^{(l')} = \sum_{m^\prime -l^\prime }^{+l^\prime } \left|
\int  Y_{l'm'}^*(\theta,\phi) \sin \theta \cos\phi
Y_{lm}(\theta,\phi)d\Omega \right|^2.
\end{equation}
When taking the imaginary part in Eq.(\ref{LAST}), only the function
\begin{math} h^{(1)}_{l'}(kr) \end{math} is complex\cite{abramowitz};
\begin{equation}
{Im}\,\left\{ih^{(1)}_{l'}(kr)\right\} \to
j_l(kr)=\sqrt{\pi/(2kr)}J_{l+1/2}(kr).
\end{equation}
Therefore, the noise spectral function
as \begin{math} \omega \to \infty  \end{math}
reads
\begin{equation}
S_{nlm}^{(xx)}(\omega) \to  \left({e^2 \over 2\omega_0 a_0^2}\right)
\sum_{l^\prime =0}^{\infty} C_{lm}^{(l')}\, \left|\int_0^\infty
r^{5/2} \chi_{nl}(r) J_{l^\prime +1/2}(kr)dr\right|^2.
\label{quasifinal}
\end{equation}
The radial integral in the above expression can be evaluated in the
limit of high frequencies and is found to vanish as
\begin{math} (ka_0)^{-(4+l+1/2)}\end{math}.
Indeed, using the hydrogen wave functions reported in the appendix,
one finds
\begin{eqnarray}
I_{nll'} &=& \int_0^\infty \sqrt{r}\left({r\over a_0}\right)^2
\chi_{nl}(r)J_{l'+1/2}(kr)dr\\
&=&-{2\over n^2} \left({(n-l-1)!\over [(n+l)!]^3}\right)^{1/2}
{1\over (ka_0)^{(4+l-1/2)}}\left({2\over n}\right)^l \,
\int_0^\infty\, dx\,  x^{(3+l-1/2)} e^{-(x/nka_0)}
\times\nonumber\\&&\phantom{xxxxxxxxxxxxxxxxxxxxxxxxxxxxxxxxxxxxx}
 L_{n+l}^{2l+1}\left({2x/ nka_0}\right)
J_{l'+1/2}(x)\nonumber
\end{eqnarray}
When \begin{math} k \to \infty \end{math} the Laguerre polynomial
can be replaced by the constant value that it takes  for a vanishing
argument \begin{math} L_{n+l}^{2l+1}(0)\end{math}. The remaining
integral is tabulated (see Eq.(6.621) in~\cite{Grad}) and one finds
\begin{eqnarray}
I_{nll'} &\approx& -{2\over n^2}
\left({(n-l-1)!\over [(n+l)!]^3}\right)^{1/2}
{1\over (ka_0)^{(4+l-1/2)}}
 \left({2\over n}\right)^l
L_{n+l}^{2l+1}(0)\Gamma(l+l'+4)
\times \nonumber \\
&&\phantom{xxxxxxxxxxxxxxxxxxxxxxx}
\left\{P^{-(l'+1/2)}_{3+l-1/2}(0) + {1\over nka_0}
\left[{d\over dx}\, P_{3+l-1/2}^{-(l'+1/2)}(0)\right]%_{x=0}
\right\},
\end{eqnarray}
where \begin{math} P_\mu^\nu(x) \end{math} are the associated
Legendre functions of the first kind. One should note that for the
sum in Eq.(\ref{quasifinal}) only few terms
are non-vanishing. This is related to well known electric dipole
selection rules which apply when calculating the quantities
\[
C_{lm}^{(l^\prime)} = \sum_{m^\prime =-l^\prime }^{l^\prime }
|\langle l^\prime m^\prime | \sin\theta \cos\phi |lm\rangle|^2,
\]
i.e.
\begin{equation}
\left< l^\prime m^\prime \right|
\sin\theta \cos\phi \left|lm\right> \neq  0
\ {\rm only\ if}
\left\{
\begin{array}{l}
m'=m\pm 1\\
l' = l\pm 1
\end{array}
\right.
\end{equation}
Applying the selection rule,
\begin{math} l^\prime =l \pm 1\end{math}, it turns out that
\begin{math}P_{3+l-1/2}^{-(l^\prime +1/2)}(0)\end{math}
vanishes for any \begin{math}l\end{math}, while its first
derivative at zero is always finite (see Eqs.(8.6.1) and (8.6.3)
in \cite{abramowitz}). This completes the proof of the asymptotic
behavior of the noise spectral functions in {\sf QM}. Taking into account
that \begin{math} ka_0 = \sqrt{2\omega/\omega_0}\end{math},
one finally concludes that for \begin{math} \omega \to \infty  \end{math},
\begin{equation}
S_{nlm}^{{\sf QM}}(\omega) \to
\left({e^2 a_{0}^2\over \omega_{0}}\right){\cal C}_{nlm}^\prime
\left({\omega_{0}\over \omega}\right)^{4+l+1/2},
\label{asyQM}
\end{equation}
where
\begin{equation}
{\cal C}_{nlm}^{'}= {2\over n^4}\sum_{l^\prime =l\pm 1}
C_{lm}^{(l')}{(n-l-1)!\over [(n+l)!]^3}[L_{n+l}^{2l+1}(0)]^2\,
\left\{
\frac{\Gamma(l+l'+4)}{n(\sqrt{2})^{4+l+1/2}}\,
\left[\frac{d}{dx}\, P_{3+l-1/2}^{-(l'+1/2)}(0)\right]%_{x=0}
\right\}^2.
\end{equation}

When comparing Eq.(\ref{asyQM}) with Eq.(\ref{asySQM}) a difference
in the two descriptions, {\sf SQM} and {\sf QM} is made {\em very
clear}. As
\begin{math} \omega \to \infty  \end{math},
{\sf SQM} predicts for noise spectral functions in state
\begin{math} \left|nlm\right> \end{math}
to decrease at large frequencies as
\begin{math}\omega^{-(3+m)}\end{math} and {\sf QM} as
\begin{math}\omega^{-(4+l+1/2)}\end{math}.

\section{Sum rules: Moments of the noise spectral functions}

In highlighting possible differences between the predictions of
Quantum Mechanics and the Stochastic Quantum Mechanics, it proves
useful to study some global properties of the spectral function,
e.g. sum rules. The zeroth order moment is readily evaluated,
\begin{eqnarray}
\label{m0}
\gamma^{(0)} & = & \nonumber
\int_{-\infty}^{+\infty} S^{A,A}(\omega) d\omega \\
 & = &\int_{-\infty}^{+\infty}  \int_{-\infty}^{+\infty}
 e^{i \omega t} G^{A,A}(t){d\omega dt\over 2 \pi}
\nonumber\\ &=& G^{A,A}(0) =
\left< A^2(0) \right>.
\end{eqnarray}
The zeroth moment \begin{math} \gamma^{(0)} \end{math} is nothing
more than the average value of \begin{math} \left<A^2\right>
\end{math} at time  zero and it is the same in {\sf QM} and {\sf
SQM}. For the special case under consideration this assumes the
value \begin{math}\left<x^2(0)\right>_{nlm}\end{math}.
\begin{equation}
\gamma^{(0)}_{nlm} =
{a_{0}^2\over 4} n^2 [5n^2+1-3l(l+1)] \,
\left[ 1 - \frac{ (l+m+1)(l-m+1)}{(2l+1)(2l+3)}
 -  \frac{ (l+m)(l-m)}{(2l+1)(2l-1)}\right].
\label{mocom}
\end{equation}
On the contrary, the calculation of the second order moment
\begin{math} \gamma^{(2)} \end{math} (again in both theories)
shows the first discrepancy between {\sf QM} and {\sf SQM}. The
computation of the second order moment can be related to the second
derivative of the two-time correlation function calculated at
\begin{math} \tau = t-t' = 0 \end{math}. Indeed repeated integration
by parts yields
\begin{equation}
\omega^2 S_N(\omega) =  - \int_{-\infty}^{+\infty}
e^{i\omega \tau }{d^2\Phi_N(\tau )\over d\tau^2 }
\left({d\tau \over 2\pi }\right).
\end{equation}
Thus
\begin{equation}
\gamma^{(2)}_N =
\int_{-\infty}^{+\infty} \omega^2 S_N(\omega)d\omega =
- \left({d^2\Phi_N(\tau )\over d\tau^2 }\right)_{\tau =0}.
\end{equation}

\subsection{Stochastic Quantum Mechanics}
In the Stochastic Quantum Mechanics from Eqs.(\ref{SQMcorr}) and
(\ref{solution}) one has
\begin{equation}
- {d^2\Phi_N(\tau =0)\over d\tau^2 }
= \left({m\hbar e\over \mu}\right)^2
\left<nlm \right|{\cos^2\phi \over r^2\sin^2\theta}\left|nlm\right>
\end{equation}
so that
\begin{equation}
\label{SQM2nd} \gamma_{({\sf SQM})nlm}^{(2)}= {e^4 \over 2 \mu
a_{0}}\left({m \over n^3}\right).
\end{equation}
In {\sf QM} one obtains
\begin{equation}
- {d^2\Phi_{nlm}(\tau =0) \over d\tau^2}=
{e^2\over 2}\left<nlm\right|\{{\ddot x}(0),x(0)\}
\left|nlm \right> .
\end{equation}
Using the Coulomb Hamiltonian
\begin{math}H= ( \bm{p}^2/2\mu )- (e^2/r)\end{math}
one may deduce the following commutators:
\begin{eqnarray}
\dot{x} &=&{i\over \hbar}\left[ H,x \right]={p_{x}\over \mu },
\nonumber\\
\ddot{x}&=&{i\over \hbar}\left[ H,p_x \right] =
- \left({e^2\over \mu}\right){x\over r^3},
\end{eqnarray}
so that
\begin{equation}
- {d^2\Phi_{nlm}(\tau =0) \over d\tau^2 }=
- \frac{e^4}{2\mu} \left< nlm \right|{x^2\over r^3}\left|nlm \right>
\end{equation}
and
\begin{equation}
\label{QM2nd}
\gamma_{({\sf QM})nlm}^{(2)}
=\left({e^4\over \mu a_{0}}{1\over n^2}\right)
\left[ 1 - {(l+m+1)(l-m+1)\over (2l+1)(2l+3)}
- {(l+m)(l-m)\over (2l+1)(2l-1)}\right],
\end{equation}
which is quite different from the expression obtained in the Bohm
theory Eq.(\ref{SQM2nd}). Clearly the differences found in the
second order moment imply rigorously different spectral functions.

\subsection{Semi-classical limit}
It is interesting to see how the two quantities which are different
for general quantum numbers have the same
semi-classical limit for large values of the quantum numbers
\begin{math} n \end{math}, \begin{math} l \end{math} and
\begin{math} m \end{math}. Setting the maximum orbital momentum
\begin{math} l=m=n-1 \end{math} one finds that
\begin{eqnarray}
\gamma^{(2)}_{({\sf SQM})n,n-1,n-1}&=&
\left({e^4\over 2 a_{0} \mu }\right)  {(n-1)\over n^3}\ , \\
\gamma^{(2)}_{({\sf QM})n,n-1,n-1}&=&
\left({e^4\over 2 a_{0} \mu }\right) {1\over n(n+1/2)}\ .
\end{eqnarray}
In the limit of  large values of $n$ the two theories agree; i.e.
\begin{equation}
\lim_{n \rightarrow \infty}\{n^2\gamma^{(2)}_{n,n-1,n-1}\}
=\left(\frac{e^4}{2 a_{0} \mu }\right)
\end{equation}
for both {\sf QM} and {\sf SQM}  as expected.

\section{Conclusions}

In this work the predictions of the Stochastic and conventinal
Quantum Mechanics have been compared in some detail with respect to
the {\em two-time correlation functions}. The example of interest in
this work is the dynamic evolution of the electric dipole moment
within the hydrogen atom. Previous numerical
computations\cite{reding} of the {\sf SQM} spectral function for the
first excited state
\begin{math} \left|211\right> \end{math}
have been confirmed.  In addition closed expressions have been
obtained in terms of modified Bessel functions \begin{math} K_n (z)
\end{math}, for several excited states  with special combinations of
quantum numbers. A derivation is provided of the asymptotic form (as
\begin{math} \omega \to \infty  \end{math}) of the noise spectral
functions for both {\sf QM} and {\sf SQM}. For large frequencies the
two descriptions provide different power law behavior. For the
hydrogen atom bound states \begin{math}\left|nlm\right>\end{math}
the {\sf SQM} spectral functions scale
\begin{math} \propto \omega^{-(3+m)}\end{math}
as opposed to the {\sf QM} spectral functions which scale
\begin{math}\propto \omega^{-(4+l+1/2)}\end{math}.

The difference in the noise spectral functions is reflected
in different sum rules which are obeyed by the spectral functions.
Explicit {\em exact} evaluations of the
second order moment of the spectral functions
\begin{math} \gamma^{(2)} \end{math},
as in Eq.(\ref{moments}), show that the two descriptions predict
indeed different values.

Finally by considering, for example, the interaction of an hydrogen
atom with the field of an electromagnetic wave, it is possible to
relate the noise spectral functions within an excited state to a
total absorption cross-section by the relation $
\sigma^N_{\text{tot}}(\omega) = 8\pi^2 \alpha_{_{\text{\scriptsize
QED}}}\, \left[\omega\, S_N(\omega)\right] $ as shown in detail in
the appendix C.

Let us briefly comment on the different predictions just so pointed
out between conventional quantum mechanics and trajectory based
interpretations of stochastic nature (Bhom or Nelson). It is
certainly worthwhile to mention here a recent work~\cite{hall} where
the author shows that trajectory based interpretations of quantum
mechanics are \emph{incomplete}. This happens for systems with
unbounded Hamiltonians. In particular it is shown that for
particular systems (providing explicit examples) there exist states
of finite energy for which the decomposition of the Schr\"{o}dinger
equation into a continuity and modified Hamilton-Jacobi equation is
impossible. These examples are also shown to be connected to the
fact that the corresponding state wave functions exhibit
\emph{fractal properties}. The main conclusion of ref.~\cite{hall}
is that \emph{Quantum Mechanics goes where trajectory
interpretations do not follow despite their (in principle) duty to
do so}. So one might wonder whether the results found in the present
work regarding two-time correlation functions are to be ascribed to
such \emph{incompleteness} of stochastic approaches to quantum
mechanics. We can just remark that the states considered here to
evidence differences in the predictions of the two theories are the
bound states eigensolutions of the Hydrogen atom hamiltonian
($|\psi_{nlm}\rangle$) and as such do not have the properties
required in~\cite{hall} to highlight the supposed incompleteness of
Stochastic QM (i.e. undefined $H\psi$, but with finite average
energy). It would certainly be interesting to consider the
possibility to construct such states for the Hydrogen atom but this
deserves further investigation, and goes beyond the scope of the
present work. Were it possible to confirm this connection it would
leave little doubt on the authors' minds as to which of the two
theories would have to be ruled out.

\begin{acknowledgments}
The authors would like to thank the referee for bringing to their
attention the interesting work of M.~J.~W Hall~{\cite{hall}} about unbounded
hamiltonians.

A.W. would like to thank the Dipartimento di Fisica,
Universit\`a di Perugia and I.N.F.N. Sezione di Perugia
for hospitality and support while this work was in progress.
\end{acknowledgments}

\appendix

\section{The Bohm description of the Hydrogen atom}

A stationary eigenstate of the Schr\"odinger equation for the
hydrogen atom is written as
\begin{equation}
\psi_{nlm}(\bm{r},t) = \chi_{nl}(r) Y_{lm}(\theta,\phi) e^{-iE_n
t/\hbar } \label{hawf}
\end{equation}
where
\begin{eqnarray}
\chi_{nl}(r) &=& -\left({2\over n^2}\right)
\left[{(n-l-1)!\over a_0^3[(n+l)!]^3}\right]^{1/2}
\negthinspace\negthinspace\negthinspace\rho^l
e^{-\rho/2}L_{n+l}^{2l+1}(\rho), \nonumber \\
\rho &=& (2r/na_0), \nonumber \\
Y_{lm}(\theta,\phi) &=& N_{lm}
P_{l}^{|m|}(\cos\theta )e^{im\phi },
\end{eqnarray}
 \begin{math}L_{n+l}^{2l+1}(\rho)\end{math}
are the associated Laguerre polynomials
and \begin{math} P_{l}^{|m|}(\cos\theta) \end{math}
are the associated Legendre functions\cite{abramowitz}.
Eq.(\ref{hawf}) can then be written as
\begin{equation}
\psi_{nlm}(\bm{r},t) = N_{lm}\chi_{nl}(r)P_{l}^{|m|}(\cos\theta)
e^{(i/\hbar )(\hbar m\phi-E_nt)},
\end{equation}
where \begin{math} N_{lm}  \end{math},
\begin{math}\chi_{n,l}(r)\end{math} and
\begin{math} P_{l}^{|m|}(\cos\theta) \end{math}
are real. Thus, the dynamics of the system in the Bohm
description is provided by the quantum action
\begin{equation}
S(r,\theta,\phi,t)= \hbar m\phi -E t.
\end{equation}
The Eqs.(\ref{eqmoto}) of motion are\cite{debro}
\begin{eqnarray}
v_{r} &=& \dot{r} ={1\over \mu}
\left({\partial S \over \partial r}\right)= 0, \nonumber \\
v_{\theta}&=& r \dot{\theta} = {1\over \mu r}
\left({\partial S\over \partial \theta}\right)=0, \nonumber \\
v_{\phi}  &=&
{1\over \mu r \sin \theta}
\left({\partial S\over \partial \phi}\right)=
r \sin( \theta ) \dot{\phi}\nonumber \\
& = & {m \hbar \over \mu r \sin \theta}\ .
\end{eqnarray}

These can be integrated yielding
\begin{equation}
r(t)=r_{0},\ \ \ \theta(t)=\theta_{0} \nonumber
\end{equation}
and
\begin{equation}
\phi(t)=\phi_{0}+
\left({m \hbar t\over \mu r^2_{0} \sin^2 \theta_{0}}\right).
\label{solution}
\end{equation}

\section{Second order moment in the Stochastic Quantum Mechanics}

Let us here consider the calculation of the moments in {\sf SQM}.
According to the definition given in Eq.(\ref{moments}) for the
state
\begin{math} \left| 211 \right> \end{math} (and for even
\begin{math} n \end{math}) one has
\begin{equation}
\gamma_{\sf SQM}^{(n)} = \left({e^2a_0^2\over 64 \omega_0}\right)
\int_0^\infty \omega^n \left({\omega_0\over \omega}\right)^{9/2}
K_1\left(\sqrt{\omega_0\over \omega}\right)d\omega .
\end{equation}
Using the change of variable
\begin{math} x = \sqrt{\omega_0/\omega}\end{math},
one finds
\begin{equation}
\gamma_{\sf SQM}^{(n)} = \left({e^2a_0^2\over 32}\right)
\omega_0^{n} \int_0^\infty x^{(6 -2n)} K_1(x) dx.
\end{equation}
This yields
\begin{eqnarray}
\gamma_{\sf SQM}^{(0)} &=& 12(e a_0)^2 \cr \gamma_{\sf SQM}^{(2)}
&=& \left({(e a_0\omega_0)^2\over 16}\right) = \left({e^4 \over 16
\mu a_0}\right).
\end{eqnarray}

\section{Hydrogen atom interacting with the field of a plane
electromagnetic wave}

Let $H_0$ denote the Hydrogen atom hamiltonian and suppose to have
an atom
 in one of his stationary eigen-states ($|N\rangle$) at $t= t_0$
\begin{equation}
|\Psi(t_0) \rangle = |N\rangle \label{initialcondition}
\end{equation}
interacting with the electric field of a plane wave of frequency
$\omega $:
\begin{eqnarray}
V_{int} (t) &=& \bm{E}(t) \cdot \bm{d}\cr \bm{E}(t) &=& E_0
\bm{\epsilon}_x \cos (kz -\omega t) \label{dipoleint}
\end{eqnarray}
the elctric field component of the plane wave assuming that it is
traveling in the z direction  with momentum $ k=\omega/c$ and
$\bm{d} = e\, \bm{x}$ is the dipole operator.

The full hamiltonian is therefore:
\begin{equation}
H =H_0 +  V_{int} (t)
\end{equation}

The quantity we would like to study is the total transition
probability per unit time given that the system is initially in the
state $|N\rangle$ at time $t=t_0$. Let us  compute first the total
transition probability $W^{\text{transition}}_N(t) $ at time $t$,
given the initial condition in Eq.~\ref{initialcondition}. At time
$t$ the system will be in the state $|\Psi(t)\rangle$ obtained from
the state $|N\rangle$ by application of the evolution operator (i.e.
solving the Schr\"odinger equation for $\Psi$). Thus the probability
$P_N(t)$ that at time $t$ the atom is still in the state $|N\rangle$
is given by:
\begin{equation}
P_N(t) = |\, \langle N | \Psi(t) \rangle \, |^2
\end{equation}
Conservation of probabily requires that:
\begin{equation}
P_N(t) + W^{\text{(transition)}}_N(t) = 1 \label{conservation}
\end{equation}
In this problem the interaction has an explicit time dependence and
so it is useful to resort to the interaction representation:
\begin{eqnarray}
|\Psi(t)\rangle &=& U(t,t_0) |\Psi(t_0)\rangle \cr U(t,t_0) &=& T
e^{+\frac{i}{\hbar}\int_{t_0}^t dt' V_{int}'(t')}\cr V_{int}'(t') &
= & e^{+\frac{i}{\hbar}H_0 t'} V_{int}(t')e^{-\frac{i}{\hbar}H_0 t'}
\end{eqnarray}
where T stands for time ordering.

Thus the probability of being in the state $|N\rangle$ at time t is
given by:
\begin{equation}
P_N(t) = | \, \langle N | U(t,t_0)\,  |N\rangle |^2
\end{equation}
{\em In second order perturbation theory} the amplitude of remaining
in the state $|N\rangle$ at time $t$ is:
\begin{equation}
 \, \langle N | U(t,t_0)\,  |N\rangle = 1 -
\frac{i}{\hbar}\int_{t_0}^t dt' \langle N |V_{int}'(t') |N\rangle -
\frac{1}{2\hbar^2} \, \int_{t_0}^t \int_{t_0}^t dt' dt'' \langle N |
T\left [V_{int}'(t') V_{int}''(t'')\right] |N\rangle
\end{equation}
The dipole interaction we are considering does not contribute at
first order since $\langle N | \bm{d}|N\rangle =0 $ for any
eigenstate of the hamiltonian $H_0$ (hydrogen atom). Thus:
\begin{eqnarray}
\langle N | U(t,t_0)\,  |N\rangle &=& 1 - \Sigma_N \cr \Sigma_N &=&
\frac{1}{2\hbar^2} \, \int_{t_0}^t \int_{t_0}^t dt' dt'' \langle N |
T\left [V_{int}'(t') V_{int}''(t'')\right] |N\rangle\cr P_N(t) &=&
1- 2 \Re( \Sigma_N )
\end{eqnarray}
Comparing this last Eq. with Eq.~\ref{conservation} one finds for
the total transition probability  $ W^{\text{transition}}_N (t) =
2\, \Re (\Sigma_N)  = 2 Im\, ( i\Sigma_N )$ or:
\begin{equation}
W^{\text{transition}}_N(t) = Im\, \left\{ + \frac{i}{\hbar^2} \,
\int_{t_0}^t \int_{t_0}^t dt' dt'' \langle N | T\left [V_{int}'(t')
V_{int}''(t'')\right] |N\rangle\right \} \label{starting}
\end{equation}
Now let us insert the explicit form of the dipole interaction given
in Eq.~\ref{dipoleint}. The fact that the wave is assumed to be
$x$-polarized selects the $x$ component of the dipole operator
$\bm{\epsilon}_x \cdot \bm{d} = d_x = e x $ and:
\begin{eqnarray}
V_{int}'(t) &=& e E_0 \, x(t)\, \cos(kz -\omega t)\cr x(t) &=&
e^{+\frac{i}{\hbar}H_0 t} x(0) e^{-\frac{i}{\hbar}H_0 t}
\end{eqnarray}
Inserting the above expression of $V_{int}(t)$ in Eq.~\ref{starting}
we also adopt the so called {\em long wavelenght approximation}
which consists in neglecting the $z-$ dependence in the interaction.
This is justified so long as
 $ka_0 \ll 1$ being $a_0$  the Bohr radius. Thus one gets:
\begin{equation}
W^{\text{transition}}_N(t) = Im\, \left\{ + \frac{i}{\hbar^2} \,
(eE_0)^2 \,
 \int_{t_0}^t \int_{t_0}^t dt' dt'' \,
\langle N | T\left [x(t') x(t'')\right] |N\rangle\,  \cos(\omega t')
\cos(\omega t'')\right \}
\end{equation}
Then note that defining $\tau = t' -t''$ and $t_+= t' + t''$ one
has:
\begin{eqnarray}
\cos(\omega t') \cos(\omega t'') &=& \frac{1}{2}\left[
\cos(\omega\tau) +\cos(\omega t_+)\right]\cr \langle N | T\left
[x(t') x(t'')\right] |N\rangle &=& \langle N | T_\tau \left [x(\tau)
x(0)\right] |N\rangle
\end{eqnarray}
where $T_\tau$ stands for time ordering relative to the $\tau$
variable. Then define $t_0= -T/2$  and $t= +T/2$,  change the
integration variables according to $dt'dt''= (1/2) d\tau dt_+ $ to
obtain:
\begin{equation}
W^{\text{transition}}_N(T) = Im\, \left\{ + \frac{i}{\hbar^2} \,
\frac{(eE_0)^2}{4} \,
 \int_{-T}^{+T} \int_{-T}^{+T}
d\tau dt_+ \, \langle N | T\left [x(\tau) x(0)\right] |N\rangle\,
\left[\cos(\omega \tau) +\cos(\omega t_+)\right]\right \}
\end{equation}
Now the time integration over the variable $t_+$ is  readily done:
\begin{equation}
W^{\text{transition}}_N(T) = Im\, \left\{ + \frac{i}{\hbar^2} \,
\frac{(eE_0)^2}{4} \,
 \int_{-T}^{+T}
d\tau \, \langle N | T_\tau\left [x(\tau) x(0)\right] |N\rangle\,
\left[2 T \, \cos(\omega \tau)
+\frac{2}{\omega}\sin\left(\frac{\omega T}{2}\right)\right]\right \}
\end{equation}
And the total transition probability per unit time
$w^{\text{transition}}_N $ (in the limit of infinite times) is
extracted:
\begin{equation}
w^{\text{transition}}_N = \lim_{T\to \infty}
\frac{W^{\text{transition}}_N(T)}{T} = Im\, \left\{ +
\frac{i}{\hbar^2} \, \frac{(eE_0)^2}{2} \,
 \int_{-\infty}^{+\infty}  d\tau \,
\langle N | T_\tau\left [x(\tau) x(0)\right] |N\rangle\, \cos(\omega
\tau) \right \}
\end{equation}
Finally it is easily shown that:
\begin{equation}
\langle N | T_\tau\left [x(-\tau) x(0)\right] |N\rangle = \langle N |
T_tau\left [x(\tau) x(0)\right] |N\rangle
\end{equation}
and hence:
\begin{equation}
w^{\text{transition}}_N = Im\, \left\{ + \frac{i}{\hbar^2} \,
\frac{(eE_0)^2}{2} \,
 \int_{-\infty}^{+\infty}  d\tau \, e^{+i\omega \tau} \,
\langle N | T_\tau\left [x(\tau) x(0)\right] |N\rangle  \right \}
\end{equation}
Thus defining the time ordered ``propagator'' by:
\begin{equation}
\alpha_N(\omega) = \frac{i}{\hbar} \, \int_{-\infty}^{+\infty}
d\tau \, e^{+i\omega \tau} \, \langle N | T_\tau\left [x(\tau)
x(0)\right] |N\rangle
\end{equation}
one writes the total transition probability per unit time as:
\begin{equation}
w^{\text{transition}}_N =\frac{(eE_0)^2}{2\hbar} \, Im\,
[\alpha_N(\omega)]
\end{equation}
This total transition probability when normalized to the  flux of
incident photons defines a total transition cross-section
$\sigma^N_{\text{tot}}(\omega)$ (adsorption and possibly stimulated
emission, if the state $|N\rangle$ is an excited state):
\begin{equation}
\sigma^N_{\text{tot}}(\omega) = \frac{8\pi\hbar\omega}{cE_0^2}\,
{w^{\text{transition}}_N}
\end{equation}
or :
\begin{equation}
\sigma^N_{\text{tot}}(\omega) = 8\pi\frac{\omega}{c}\,
\frac{e^2}{2}\,Im \,[\alpha_N(\omega)]
\end{equation}
On the other end  we have shown, c.f. Eq.~\ref{connection} that the
imaginary part of $\alpha_N(\omega)$ is directly related to the
spectral function (fourier transform) of the two-time correlation
functions:
\begin{equation}
\frac{\hbar}{2\pi} Im\, [\alpha_N(\omega)] = S_N(\omega) =
\int_{-\infty}^{+\infty} d\tau \,e^{+i\omega\tau}\, \frac{1}{2}\,
\langle N | [x(\tau)x(0) + x(0)x(\tau)] |N\rangle
\end{equation}
We therefore conclude:
\begin{equation}
\sigma^N_{\text{tot}}(\omega) = 8\pi^2 \alpha_{_{\text{\scriptsize
QED}}}\, \left[\omega\, S_N(\omega)\right]
\end{equation}
$\alpha_{_{\text{\scriptsize QED}}} = e^2/(\hbar c) \approx 1/137$
being the fine structure constant.

\end{document}